\documentclass[journal]{IEEEtran}
\usepackage{cite}
\usepackage{amsmath,amssymb,amsfonts}
\usepackage{algorithmic}
\usepackage{graphicx}
\usepackage{textcomp}
\usepackage{wrapfig}

\ifCLASSINFOpdf
\else
\fi
\begin{document}
\title{American Sign Language Recognition \\ Using RF Sensing} 
\author{Sevgi~Z.~Gurbuz,~\IEEEmembership{Senior Member,~IEEE}, Ali~C.~Gurbuz,~\IEEEmembership{Senior Member,~IEEE},
Evie A. Malaia, Darrin J. Griffin, Chris Crawford, M. Mahbubur Rahman, Emre~Kurtoglu, Ridvan Aksu, Trevor~Macks and Robiulhossain Mdrafi, ~\IEEEmembership{Student Member,~IEEE}

\thanks{S.Z. Gurbuz, T. Macks, M.M. Rahman, E. Kurtoglu, and R. Aksu are with the Department
of Electrical and Computer Engineering, University of Alabama, Tuscaloosa,
AL, 35487 USA (e-mail: szgurbuz@ua.edu, tfmacks@crimson.ua.edu, mrahman17@crimson.ua.edu, ekurtoglu@crimson.ua.edu, raksu@ua.edu).}
\thanks{A.C. Gurbuz and R. Mdrafi are with the Department
of Electrical and Computer Engineering, Mississippi State University, Starkville,
MS, 39762 USA (e-mail: gurbuz@ece.msstate.edu and rm2232@msstate.edu).}
\thanks{D. Griffin is with the Department
of Communication Studies, University of Alabama, Tuscaloosa,
AL, 35487 USA (e-mail: djgriffin1@ua.edu).}
\thanks{E. Malaia is with the Department
of Communication Disorders, University of Alabama, Tuscaloosa,
AL, 35487 USA (e-mail: eamalaia@ua.edu).}
\thanks{C. Crawford is with the Department
of Computer Science, University of Alabama, Tuscaloosa,
AL, 35487 USA (e-mail: crawford@cs.ua.edu).}

\thanks{Manuscript accepted August 2020.}
}
\markboth{ Submitted to IEEE Sensors Journal}%
{Shell \MakeLowercase{\textit{et al.}}: Bare Demo of IEEEtran.cls for IEEE Journals}

\maketitle

\begin{abstract}
Many technologies for human-computer interaction have been designed for hearing individuals and depend upon vocalized speech, precluding users of American Sign Language (ASL) in the Deaf community from benefiting from these advancements.  While great strides have been made in ASL recognition with video or wearable gloves, the use of video in homes has raised privacy concerns, while wearable gloves severely restrict movement and infringe on daily life.  \textbf{Methods:}  This paper proposes the use of RF sensors for HCI applications serving the Deaf community.  A multi-frequency RF sensor network is used to acquire non-invasive, non-contact measurements of ASL signing irrespective of lighting conditions.  The unique patterns of motion present in the RF data due to the micro-Doppler effect are revealed using time-frequency analysis with the Short-Time Fourier Transform.  Linguistic properties of RF ASL data are investigated using machine learning (ML).  \textbf{Results:}  The information content, measured by fractal complexity, of ASL signing is shown to be greater than that of other upper body activities encountered in daily living.  This can be used to differentiate daily activities from signing, while features from RF data show that imitation signing by non-signers is 99\% differentiable from native ASL signing.  Feature-level fusion of RF sensor network data is used to achieve 72.5\% accuracy in classification of 20 native ASL signs.  \textbf{Implications:}  RF sensing can be used to study dynamic linguistic properties of ASL and design Deaf-centric smart environments for non-invasive, remote recognition of ASL.  ML algorithms should be benchmarked on native, not imitation, ASL data.
\end{abstract}


\section{Introduction}
\IEEEPARstart{U}{sers} of American Sign Language (ASL) make up over 1 million people in the U.S. and Canada, based on statistics provided by Gallaudet University (the world’s only university designed to be barrier-free for deaf and hard of hearing students located in Washington, D.C.).  People in the Deaf community, who rely on ASL as their primary mode of communication, rely heavily on technology as an assistive device as they navigate communication/language barriers that status quo society often creates.  Unfortunately, many technologies are designed for hearing individuals, where vocalized speech is the preferred mode of communication, and has driven a burgeoning market of voice recognition software and voice-controlled devices.  This precludes the Deaf community from benefiting from advances in technology, which if appropriately designed to be responsive to ASL, could in fact generate tangible improvements in their quality of life.

Although there has been much research related to technologies for the deaf or hard of hearing (HoH) over the past three decades, much of this work has focused on the translation of sign language into voice or text using camera-based or wearable devices.  Examples of wearable sensors include inertial measurement units (IMUs) containing accelerometers, gyroscopes, and magnetometers to model hand movement, electromyography (EMG) to monitor muscle contractions, and strain gauges to measure skin deformation.  Although sensor-augmented gloves \cite{Tub15,Lee18} have been reported to typically yield higher gesture recognition rates than camera-based systems \cite{Gar17,Beh17}, they cannot capture the intricacies of sign languages presented through head and body movements.  Facial expressions are effectively captured by optical sensors; however, video cameras require adequate light and a direct line-of-sight to be effective.

While the emphasis on ASL translation contributes towards facilitating interaction between hearing and Deaf individuals, it overlooks issues surrounding the way technology will impact the daily lives of Deaf individuals.  Wearables are invasive devices, which limit signer's freedom in conducting daily activities and is not designed with ASL movements and language constraints in mind, while video cameras trigger concerns over privacy and potential surveillance.  Previous investigations of prototypes for ASL translation often fail to involve participants or investigators fluent in ASL, leading to a deficiency in addressing the best way technology can be used to serve the \textit{needs} the Deaf community - a goal that is broader than translation and relates to the design of smart environments for the Deaf.  This includes, for example, the design of smart Deaf spaces  augmented with sensors that can respond to the natural language of the Deaf community for the purposes of environment control, remote health, and security.

In this context, RF sensors have several important advantages over alternative sensing modalities, which make them uniquely desirable for facilitation of human-computer interaction (HCI). RF sensors are non-contact and completely private, fully operational in the dark, and can even be used for through-the-wall sensing.  Most importantly, RF sensors can acquire a \textit{new source of information} that is inaccessible to optical sensors: visual representation of kinematic patterns of motion via the micro-Doppler ($\mu$D) signature \cite{ChenMDBook}, as well as radial velocity measurements and range profiles.  This has enabled the use of RF sensing  across a variety of applications \cite{AminBook}, including fall detection \cite{Amin_SigPro_2016}, activity recognition \cite{Gurbuz2017,LiuWiFiSurvey2019}, pedestrian detection \cite{Bilik_SigPro_2019}, gesture recognition \cite{Molchanov2015,WiGest2015,GoogleSoli2016,Latern2018,Hazra2019}, as well as heart rate, respiration, and sleep monitoring \cite{BoricLubecke2016,LiuWiFiresp2016}. ASL signs have been used as example classes in Wi-Fi based gesture recognition studies \cite{WiFinger2016}, while imitation signing was utilized in two other Wi-Fi studies \cite{WiSign2017} \cite{SignFi2018}.

This paper presents an in-depth examination of RF sensing for the recognition of ASL for Deaf-centric design of smart environments.  To the best of our knowledge, this study represents the first study of RF-based recognition of native ASL signing.  In Section II, we discuss the importance of including perspectives of those in the Deaf community through organizational partnerships and present the results of a focus group that reveals the way existing technologies for ASL translation are perceived.  Section III presents the methodology and experimental design for measurement of ASL signs using RF sensors, including distinction between native signing and imitation signing.  In Section IV, the micro-Doppler signature of ASL signs acquired from different RF sensors are discussed.
The linguistic properties of ASL that are observable via RF sensing is presented in detail in Section V.  This includes investigation of the information content of ASL versus daily gestures via fractal complexity analysis, as well as differences between native and imitation signing.  In Section VI, a variety of handcrafted features are extracted and the resulting classification accuracy compared for up to 20 ASL signs.  Key conclusions are summarized in Section VII.

\section{Partnerships \& Focus Group}
A common problem with information communicative devices (ICTs) that are developed to assist people who are deaf and hard of hearing is that the developers often know virtually nothing about deafness or Deaf culture, are not fluent in ASL, and bring a hearing perspective to design features. There have been many previous attempts to create new technologies for communication with ASL users, such as signing gloves that were designed with the intention of interpreting sign and translating it into written text requiring users to wear gloves to communicate. However, as this would be akin to wearing a mask to speak to someone using a different oral language and might be uncomfortable, technologies such as signing gloves have been rejected by and evoked negative feedback from the Deaf community for their lack of cultural and user sensitivity. 

Thus, the active involvement of community stakeholders can make important contributions not only through shared knowledge, but also by ensuring researchers understand the problems that matter the most to the community.  In the spirit of the expression \textit{``Nothing about us without us"} \cite{Chui3883,Charlton1998} we have espoused a Deaf-centric design philosphy:  our research  team includes one member who identifies as Culturally Deaf and is fluent in ASL, while faculty and staff at the Alabama Institute for Deaf and Blind (AIDB) and Gallaudet University provide cultural and linguistic information throughout the research process (from planning to execution).

Our study of RF sensing as a means for facilitating the design of smart environments was motivated by a focus conducted in collaboration with AIDB with 7 Deaf participants.
 While a detailed discussion of all responses is beyond the scope of this paper, we do feel it significant to share that many participants indicated a need for technologies that will help ease the communication issues they encounter, especially in times of emergency or with regard to security matters. Participants expressed frustration with wearable gloves that they described as “inaccurate” and invasive. Multiple focus group attendees also raised concerns about surveillance if video is used in the home (as opposed to cell phones, which they found quite helpful for communications). Furthermore, the participants  indicated a desire for Deaf-friendly personal assistants, which could aid in everyday operations such as scheduling, remotely turning on and off lights, or even making phone calls through the use of ASL.  They were also thrilled with the idea that a non-invasive smart-environment responsive to ASL could be designed.




\section{Experiment Design and RF Datasets Acquired}
Machine learning algorithms are data greedy methods that require large amounts of training samples to enable the network to learn complex models.  Thus, some researchers have resorted to acquiring data from non-native signers, who may not know any ASL, as an expeditious source of data.  Although ASL is often likened to gesturing, it is important to recognize that ASL is a \textit{language}, and not reduce signing to mechanical hand and arm movements that can be easily imitated. Thus, while gestures can be made using any participant, studies of ASL require participants for whom sign language is their native language, e.g. Deaf / Hard-of-Hearing individuals. Consequently, two distinct datasets were acquired for comparative study: 1)
\textbf{\textit{native ASL data}} from Deaf participants and 2) \textbf{\textit{imitation data}} from hearing individuals imitating ASL signs based on copy-signing videos.

\subsection{Experiment Design and Procedure}
\begin{figure*}
\centering
\includegraphics[width=17.5cm]{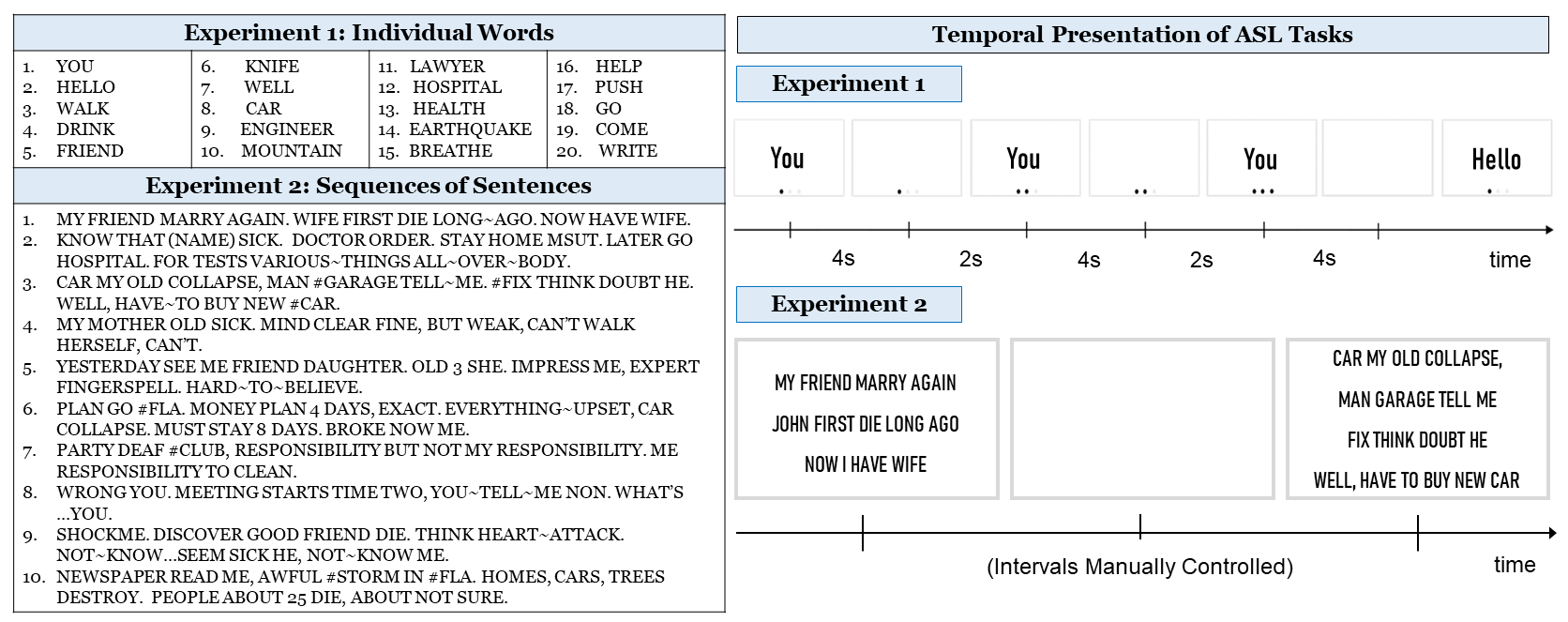}
  \caption{Description of ASL experiments: words and temporal presentations.}
  \label{study_design}
\end{figure*}

Two blocks of experiments were conducted in this study:  Experiment 1 consisted of individual words/signs, while Experiment 2 involved sequence of sentences.  The individual words were selected from  the  ASL-LEX  database \cite{ASL_LEX_2016}  (http://asl-lex.org/),  choosing words that are higher frequency, but not phonologically related to ensure a more diverse dataset. Sentences were chosen from those  used  in  previous  linguistic  studies  of  ASL  by  one  of the co-authors \cite{mcdonald2016}. Figure \ref{study_design} provides a listing of the words and sentences used in experiments.


A total of 3 Deaf participants took part in the ASL data collection, while imitation data was captured from 10 hearing participants.   In all experiments, participants were asked to begin with their hands placed on their thighs, and to return to this position once done signing. The study lasted approximately 75 minutes for each participant 
and was comprised of the following steps:

\begin{enumerate}
\item Participants entered a conference room, read, and signed an informed consent form.
\item An experimenter introduced participants to the RF sensing system and provided an overview of the study procedures during a 30-minute informative session. 
\item Participants entered a lab outfitted with RF sensors and were introduced to the test environment.  To ensure comfort with their surroundings, participants were also offered the opportunity to ask questions about RF sensing.
\item \textit{{[Imitation Study Only]}} Hearing participants practiced ASL signs during a 15 minute training session with a Child-of-Deaf Adult (CODA) fluent in ASL until they were comfortable responding to visual prompts.
\item \textit{{[Imitation Study Only]}} Hearing participants were shown a copy-signing video where a CODA enacted the desired sign, after which the participant was expected to repeat the same sign.  Participants were presented with a random ordering of single-word signs (Fig. \ref{study_design}) to foster independence in each repetition of the signs.
\item Deaf participants were prompted with text-based visuals and asked to demonstrate the ASL sign for the individual word (Fig. \ref{study_design}) shown on the monitor. Words appeared for 4s, with an inter-stimulus interval of 2s. Three repetitions of each word were collected per participant.
\item Deaf participants were next asked to sign 10 `sequences of sentences' (Fig. \ref{study_design}) in response to text-based prompts. Each sequence of sentences was collected once per participant, resulting in 3 samples per sentence.   Due to the variation of time needed to complete the sentence sequences, stimulus intervals were manually controlled. 
\item Participants engaged in semi-structured interviews designed to learn more about their experiences.
\end{enumerate}

\subsection{Test Setup of RF Network}
RF sensors operating at three different transmit frequencies are considered in this work.  The Xethru sensor is a low-power ultra-wide band (UWB) impulse radar with a transmission frequency range of 7.25 - 10.2 GHz as well as 65$^{\circ}$ azimuth and elevation beamwidth.  The range resolution of an RF sensor is given by $c/2\beta$, where $c$ is the speed of light and $\beta$ is the bandwidth.  Thus, the Xethru sensor has about 5 cm range resolution.  Frequency modulated continuous wave (FMCW) radars at 24 GHz and 77 GHz were also deployed.  The 24 GHz system by Ancortek was operated with bandwidth of 1.5 GHz, while the 77 GHz Texas Instruments device transmitted with a bandwidth of 750 MHz.  This resulted in range resolutions of 10 cm and 20 cm, respectively.  It is important to note that RF sensors can have finer resolution as devices of greater bandwidth become increasingly widespread and low cost. 

\begin{figure*}
\centering
\includegraphics[width=17.5cm]{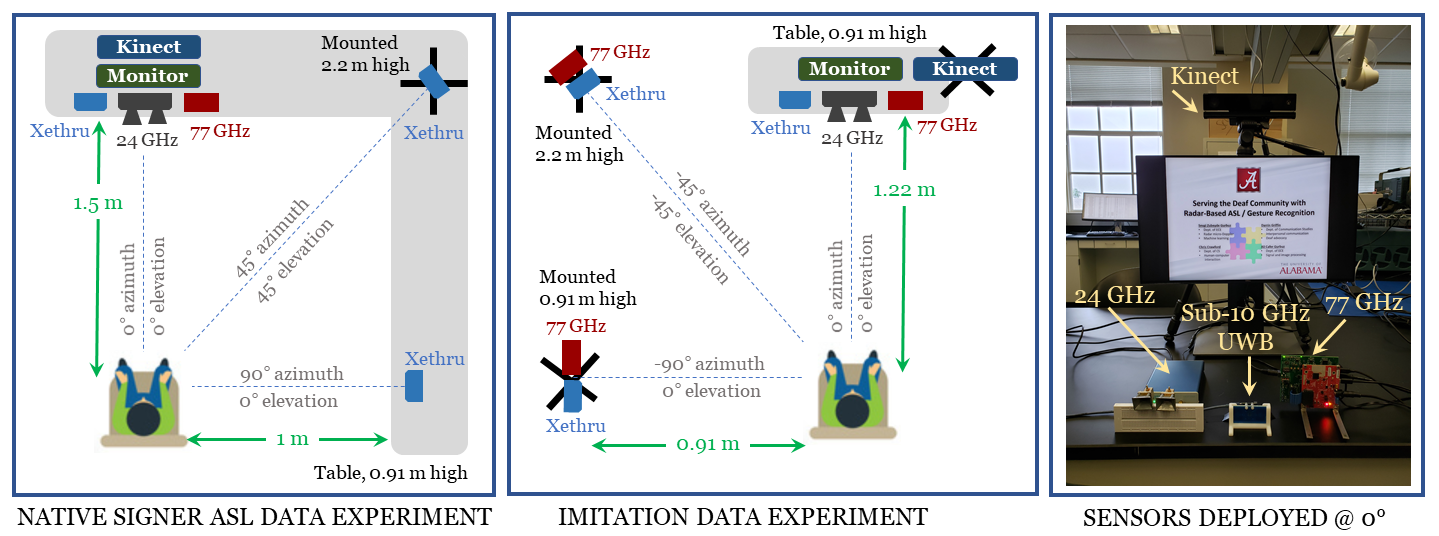}
  \caption{Test setup used for acquisition of ASL and imitation datasets.}
  \label{fig:testsetup}
\end{figure*}

Several RF sensors and a Kinect sensor were deployed at various positions to observe the participant from different perspectives, as shown in Fig. \ref{fig:testsetup}.  Participants were asked to sit on a bar stool facing a computer monitor, which was used relay prompts indicating the signs to measured. The monitor was placed just behind the RF and Kinect sensors so that the visual cues would ensure the participant remained facing forward throughout the experiment. The Kinect data were used for comparison with radar data, and to annotate the RF micro-Doppler signatures and optical flow plots given in this paper.  Annotations were performed by a CODA fluent in ASL who made manual notations of the Kinect video frames, which were then correlated to the time axis of the radar data. 

\subsection{Pre-Processing and Representation of RF Data}

The signal received by a radar is, in general, a time-delayed, frequency-shifted version of the transmitted signal.  In many practical scenarios, it has been shown that the scattering from the human body can be approximated using the superposition of returns from $K$ points on the body \cite{ChenMDBook}. Thus, 
\begin{equation}
\label{eq:rre}
x[n] = \sum_{i=1}^{K}{a_{i}exp\Bigg\{-j\frac{4\pi f_{c}}{c}R_{n,i}\Bigg\}},
\end{equation} 
where $R_{n,i}$ is the range to the $i^{th}$ body part at time $n$, $f_{c}$ is the transmit center frequency, $c$ is the speed of light, and the amplitude $a_{i}$ is the square root of the power of the received signal as given by 
\begin{equation}
a_{i} = \frac{\sqrt{G_{tx}G_{rx}} \lambda \sqrt{P_{t} \sigma_{i}}}{(4\pi)^{3/2} R_{i}^2 \sqrt{L_{s}} \sqrt{L_{a}}}
\end{equation} 
Here, $G_{tx}$ and $G_{rx}$ are the gains of the transmit and receive antennas, respectively; $\lambda$ and $P_t$ are the wavelength and power of the transmitted signal, respectively; $\sigma_{i}$ is the radar cross section (RCS) of the $i^{th}$ body part; $L_{s}$ and $L_{a}$ are system and atmospheric losses, respectively.

\subsubsection{RF Micro-Doppler Signature}

The data provided by each RF sensor is a time-stream of complex in-phase/quadrature (I/Q) data, as modeled by (\ref{eq:rre}).  The effect of kinematics is predominantly reflected in the frequency modulations of the received signal. While the distance of the target to the radar does affect the received signal power, these variations are not significant for typical people. Rather, the amplitude of the received signal is primarily dependent upon the transmit frequency being used and RCS of target observed.  Thus, at higher frequencies, such as 77 GHz, the received signal power from a person will be much lower than that of a radar at lower frequencies, such as 24 GHz or below 10 GHz.  Not just transmit frequency, but also bandwidth, pulse repetition interval (PRI), observation duration, and aspect angle have been shown to affect the radar received signal and the performance of machine learning algorithms applied to the RF data \cite{Gurbuz2015}.

To reveal patterns of motion hidden in the amplitude and frequency modulations of the received signal, time-frequency analysis is often employed. The \textit{micro-Doppler signature}, or spectrogram, is found from the square modulus of the Short-Time Fourier Transform (STFT) 
of the continuous-time input signal. It reveals the distinct patterns caused by micro-motions \cite{ChenMDBook}, e.g. rotations and vibrations, which result in micro-Doppler frequency modulations centered about the main Doppler shift related to translational movement.

\subsubsection{Removal of Clutter and Noise}
Prior to computation of the spectrogram, a 4th order high pass filter is applied to remove reflections from stationary objects, such as the walls, tables, and chairs.  The STFT itself is computed using Hanning windows with 50\% overlap to reduce sidelobes in the frequency domain and convert the 1D complex time stream into a 2D $\mu$D signature.  It is common for there still to be some noise components in the data due to electronic noise or other sensor-related artifacts. Such artifacts were particularly evident in the 24 GHz FMCW and sub-10 GHz UWB sensors, so an isodata thresholding method was applied to eliminate any background noise. The isodata algorithm \cite{Young1998FundamentalsOI} automatically finds an optimal threshold for a given image.  
Any time-frequency bin that has value less than the threshold is set to zero. 

\subsubsection{Selection of Input Dimensionality}

\begin{figure}[t]
\centering
\includegraphics[height=6.5cm]{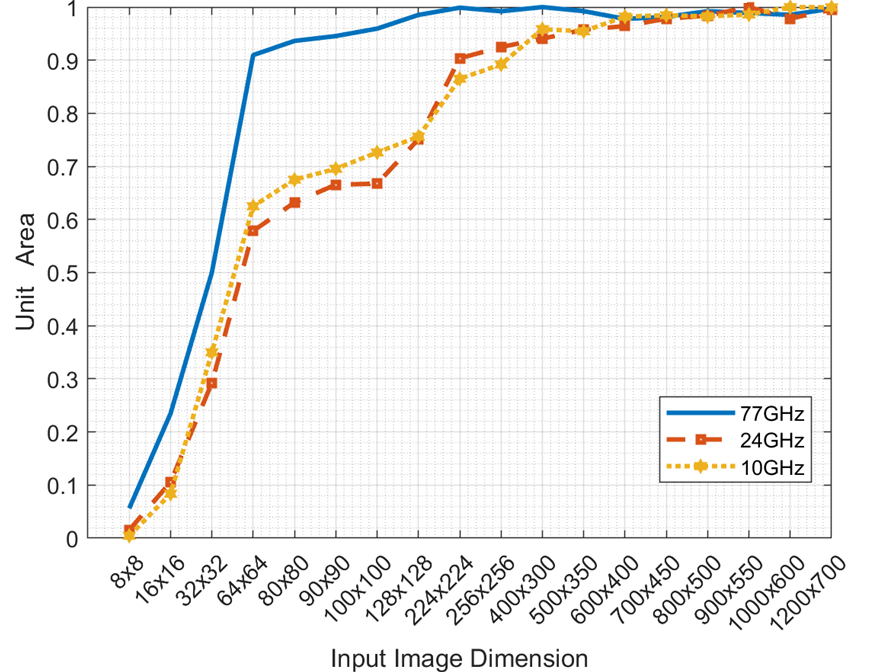}
  \caption{Surface area of feature space vs image size.}
  \label{fig:inputdim}
\end{figure}

The size of spectrogram used in this work is determined based on an analysis of change in surface area of the feature space as a function of size.  Principal Component Analysis (PCA) is applied to spectrograms of different dimensions and an n-dimensional convex hull applied to determine the boundaries of the PCA-based feature space.  
For each sensor, Figure \ref{fig:inputdim} plots the surface area of the the feature space for different input dimensions. As the input dimensions increase, the surface area of the convex hull of the feature space levels off.  The 77 GHz sensors exhibits immediate leveling beyond an input dimension of $65 \times 65$, while the 10 GHz and 24 GHz sensors more gradually level off. Based on this result, we thus reduced the size of the $\mu$D signatures for all sensors to $65 \times 65$.

\section{Visualization of RF Measurements of ASL}

\begin{figure*}
\centering
\includegraphics[height=15cm]{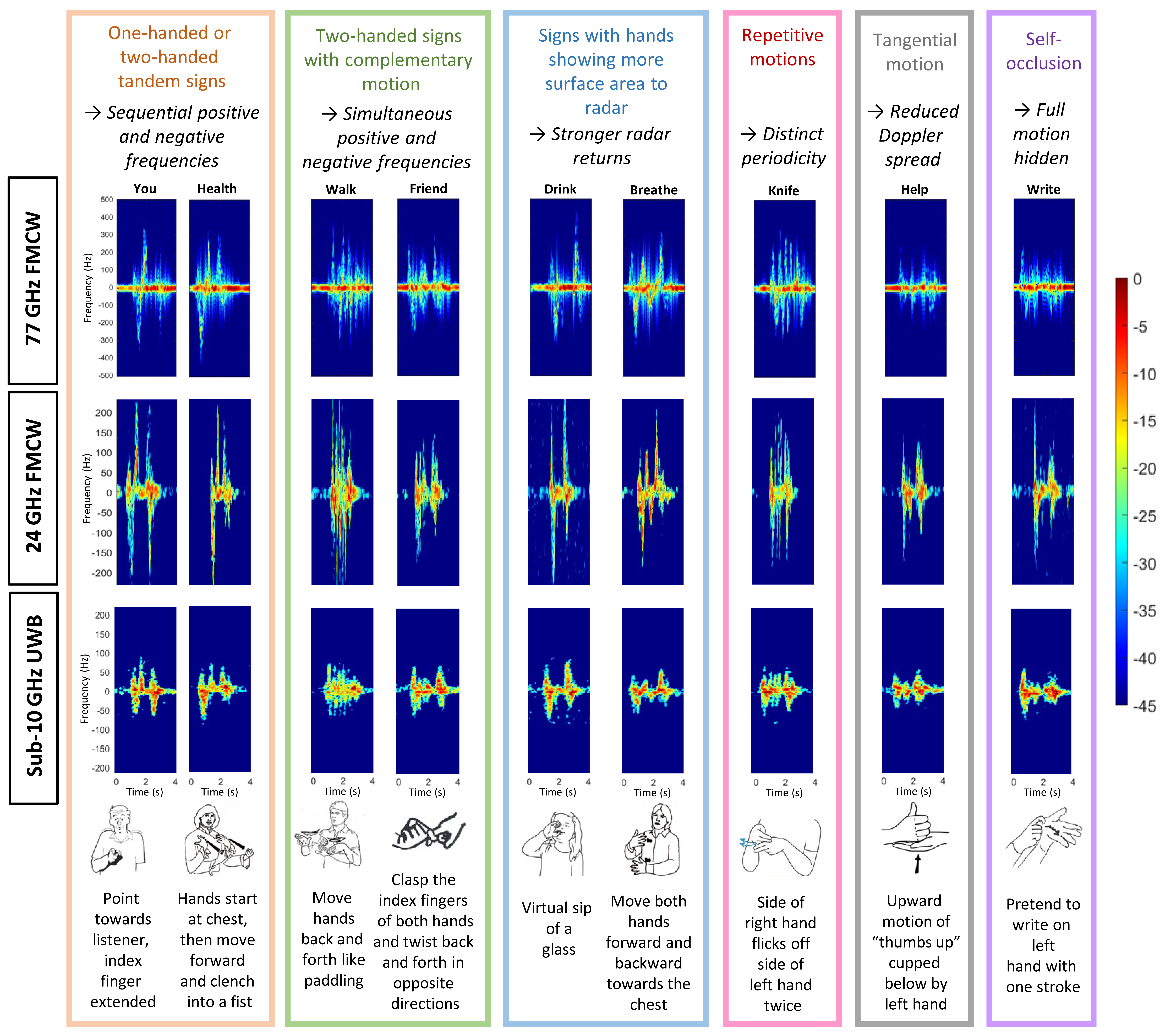}
  \caption{Micro-Doppler signatures of nine ASL signs acquired by three different RF sensors.}
  \label{fig:signs1}
\end{figure*}

Illustrative examples of the $\mu$D signatures for several ASL signs are shown in Figure \ref{fig:signs1}.  For these examples, each RF sensor is positioned so as to be directly facing the signer with a 0$^{\circ}$ aspect angle.  Comparing the {$\mu$}D signatures for different sensors, it may be observed that as the transmit frequencies increase so does the peak Doppler shift. This result is consistent with expectations as the Doppler shift resulting from the motion of an object with velocity $v$ towards a radar is $f_{D}=2v(f_{c}/c)$.  From the perspective of ASL recognition, the greater Doppler spread observed at higher frequencies also results in subtle, small micro-motions being revealed in the signature; e.g.,  the 77 GHz $\mu$D signatures appear crisper and more nuanced than those obtained with the lower frequency devices. 

Inferences about kinematics can be made from the {$\mu$}D:
\begin{itemize}
    \item \textit{The starting position of the articulators (hands) affects the initial and final frequency components measured.}  
    Raising the arms from their initial position on the thighs results in motion away from the radar, resulting in negative frequency spikes at the beginning of all samples.
    \item \textit{One-handed or tandem signing can be discerned from two-handed free signing.}  A {$\mu$}D signature having either a positive or negative Doppler frequency at a given time is indicative of the hand(s) either moving toward or away from the sensor, as in the one-handed sign \textsc{you} and tandem sign \textsc{health}.  Otherwise, both positive and negative Doppler are present, as with the signs \textsc{walk} and \textsc{friend}, which have two-handed complementary motion.
    \item \textit{When more surface area of the hand(s) faces the radar line-of-sight, the received power of the signal may be observed to be greater.} For example, the  \textsc{drink} signature has two vertical spikes, due to raising and lowering the cupped hand.   The first peak has a greater intensity as the outside of the hand faces the radar. When the hand is rotated to its side, the intensity is lessened.
    \item \textit{In signs with reduplicated movements, the number of cycles can be counted}; e.g. \textsc{knife} and  \textsc{walk}.
    \item \textit{The effect of aspect angle between the line-of-sight and direction of motion can be observed in the signatures.} For example, the sign for  \textsc{help} involves primarily vertical movement, which is orthogonal to the radar line-of-sight, and hence has a low Doppler spread. 
    \item \textit{The effects of occlusion can be observed in some signatures.} In \textsc{write}, the right hand pretends to hold a pen while swiping across the inside of the left palm, which partially shields the right hand from the radar. This causes reduced Doppler spread in relation to fully exposed motion, such as in     \textsc{knife}. 
\end{itemize}

\begin{figure}[t]
\centering
\includegraphics[height=4.4cm]{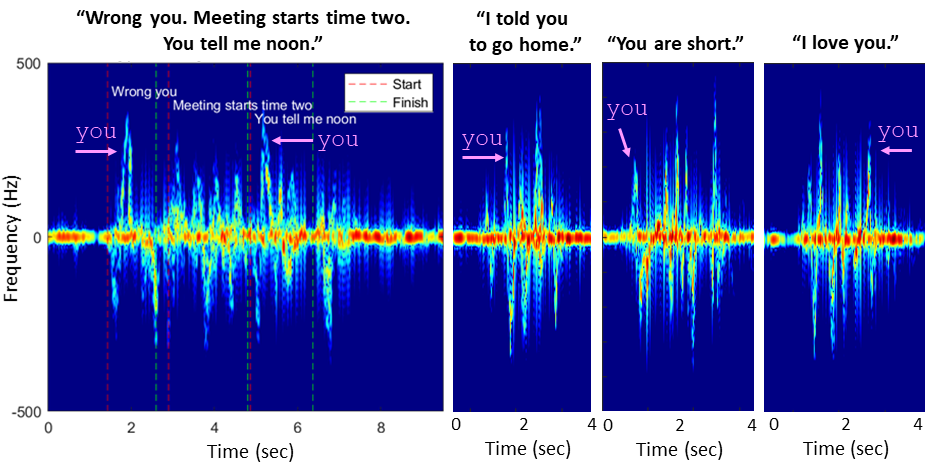}
  \caption{Coarticulation example: $\mu$D signature for \textsc{you} in different positions within a sentence.}
  \label{fig:YOUsigns}
\end{figure}
 In speech, quantitatively characterized temporal dynamics of vocal signal provided insights into mathematical properties of information exchange. Although quantification of temporal properties of signed signal is behind that of speech, we know that dynamic properties of signs (e.g. dominant hand velocity, temporal contour of motion signature for manual and non-manual articulators) contribute crucial linguistic information to the meaning of signs \cite{Evie4, Evie5, malaia2013kinematic}. Although radar data does not provide for easy identification of hand shapes and place of articulation (i.e. static spatial features of signs), it does allow for improved measurements of the gross spatiotemporal dynamics of signing (i.e. shape change dynamics), combining information picked up from the moving hands with the information on other articulators (head and body).  Remembering that ASL is a natural language, and not merely gesture, linguistic features of the RF ASL data can contribute to motion recognition, while, conversely, machine learning can also be used to identify linguistic properties.

\subsection{Coarticulation}

An important linguistic feature that is visible in the RF micro-Doppler signatures of ASL is \textit{coarticulation}: the effect of the previous sign (more specifically, its place of articulation) influencing subsequent signs.  In sign language, coarticulation affects both the total distance travelled and the speed of the articulator motion.  These kinematic features can be observed in the $\mu D$ signature and result in signs appearing slightly differently in different sentences.  
Consider, for example, the sentence \textsc{``Wrong you. Meeting starts time two. You tell me noon.''} depicted in Figure \ref{fig:YOUsigns}, which includes the sign for \textsc{you} twice.  This sign is visible in the $\mu D$ signature as the two peaks with the greatest positive $\mu D$ frequency.  The extension and retraction of the hand is reflected as two separate lines forming the sides of a narrow triangle.  However, comparing the first and second occurrences, it may be observed that in the second occurrence the sign is entered rapidly, almost instantly reaching the peak of the $\mu D$ signature, whereas the first occurrence is slower: the micro-Doppler has a finite slope, indicating that a longer duration was needed to reach peak speed.  This makes sense considering that more time is needed for the hands to travel from their initial position on the knees as opposed starting the sign in mid-air. A similar shape for the sign  \textsc{you} may be observed in the sentence \textsc{``You are short''}, when \textsc{you} also is the first sign in the sentence. 

In contrast, in mid-sentence occurrences of \textsc{you}, as in \textsc{``I told you to go home''}, the burst is so narrow that the extension and retraction can no longer be distinguished. And in the final example \textsc{``I love you.''} the sign yields greater reflected power due to the prior sign involving both arms crossed over the chest (\textsc{love}).  As the arms uncross to permit the right hand to extend forward, strong reflections are generated from the arms crossed across the chest. The two arms uncrossing to permit the right hand to extend forward results in stronger reflection signal than that observed in the previous instances of the sign \textsc{you}.

\subsection{Fractal Complexity of Signing Versus Daily Activity}

Analysis of information content in speech vs. everyday motion using the visual properties of the signal and optical flow \cite{Evie6,Evie7} has indicated that signers transmit more information (in the sense of mathematical entropy) than humans carrying out dynamic tasks, and that the intelligibility of a signing stream is crucially dependent on the ability of signers to parse entropy changes in visual information \cite{Evie8,Evie9}. Our goal in the present analysis is to build on current understanding of human signal parsing for sign language \cite{Evie7}, extending existing analytical approaches for use with data collected using RF sensors. Thus, we demonstrate that RF data can be used for distinguishing between signing and biological motion (e.g. daily activities) that occur in the house.

One way of evaluating the information transfer over time due to human motion is the \textit{fractal complexity} \cite{Evie6} of the optical flow (OF).  Optical flow is a technique often used in computer vision to determine the apparent motion of image objects between two consecutive video frames caused by either the movement of the object or camera.  Radar data is not acquired in the form of a video; however, the raw complex I/Q time stream from RF sensors with linearly frequency modulated transmissions, such as those used in this study, can be converted into a \textit{range-Doppler-time (RD) data cube} by computing the 2D FFT of the data for each sweep of the FMCW radar \cite{ErolRDC2019}.  Then, each slice in time (or frame, as termed in video processing) is a range-Doppler map.  The 3D RD cubes enable the simultaneous observation of the radial distance to the radar as well as the radial velocity, and provide an alternative to $\mu$D signatures for visualizing RF ASL data.

\begin{figure}[t]
\centering
\includegraphics[height=6.5cm]{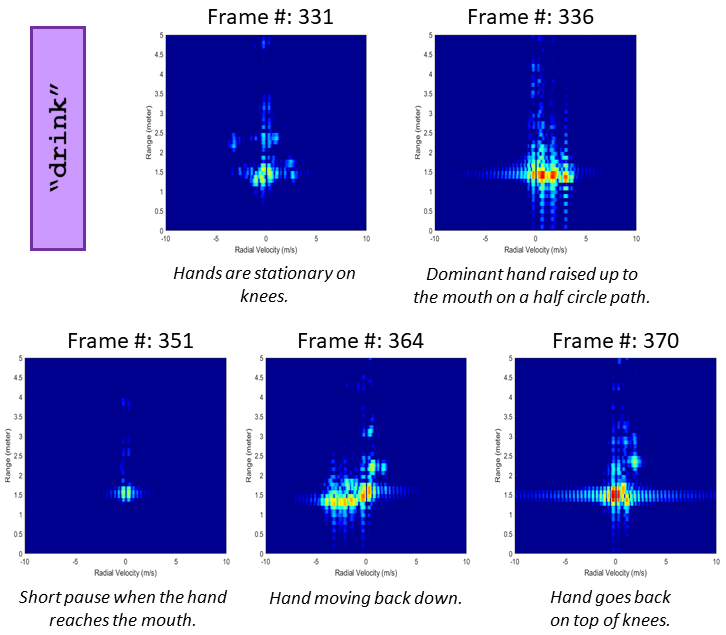}
  \caption{RD maps at various times for the sign \textsc{drink}.}
  \label{fig:RDvideo}
\end{figure}

\subsubsection{Radar Data Cube}
Consider, the RD data cube for the sign \textsc{drink}, shown in Figure \ref{fig:RDvideo} for the 77 GHz sensor.  The response from the torso is observable in the initial RD maps as strong signal returns at 0 Hz.  When the hand(s) move towards the sensor, a positive Doppler shift is incurred, with the peak speed given by maximum Doppler shift.  The distance towards the sensor can be found via number of range bins that the strong returns have shifted.  Arm motion is indicated by multiple peaks at various speeds, e.g. Frame \#336 of \textsc{drink}.  Negative Doppler occurs for motion away from the sensor.

The number of pixels that are spanned by arm and hand movements is dependent upon the bandwidth of the RF sensor.  Higher bandwidth results in a smaller distance spanned by each pixel.  IN the RD data cube of \textsc{drink}, it may be observed that the hands move about 2 range bins closer to the radar than the torso.  At a bandwidth of 750 MHz for the 77 GHz sensor, each range bin corresponds to about 20 cm displacement.  For a person with average arm length, during the course of enacting the \textsc{drink} sign, the radial displacement of the arms is about 25-30 cm.  This is consistent with the radar measurement of 2 range bins. The range resolution of the radar can be improved through transmission of a waveform with greater bandwidth, resulting in more accurate measurement of hand displacement.

\begin{figure}[t!]
\centering
\includegraphics[height=5.9cm]{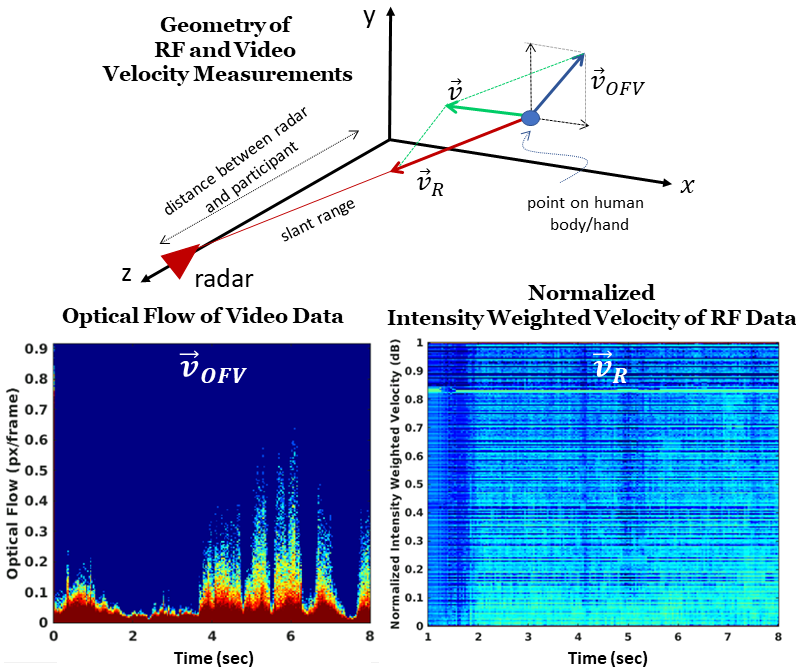}
  \caption{Comparison of video optical flow in cross plane with intensity weighted velocity diagram of 77 GHz RF sensor.}
  \label{fig:OF}
\end{figure}

\subsubsection{Intensity-Weighted Velocity (IWV) Diagram}
In video, the 2D slice at each time frame represents horizontal and vertical distance.  Thus, computing the optical flow in video reveals the distance traveled by each pixel as it moves from frame to frame (i.e., velocity), while the intensity of optical flow is proportional to area of the moving part.  

However, RF sensors measure the radial velocity ($v$), which can be found from the x-axis of a RD map using the Doppler shift ($f_{D}$) relation, $f_{D}=2 v f_{c}/c$. The intensity of each pixel in the RD map relates to the received signal power, which is proportional to the radar cross section (RCS) of the body part. RCS depends on a multitude of factors, including frequency, shape, material properties, aspect angle as well as physical dimensions.  To compute a representation comparable to video OF, the velocity corresponding to each pixel in the RD map is weighted according to its intensity and binned.  In this way, we can ensure the information of both the velocity and the area of moving objects in the same speed range, which correspond to OF magnitude and the intensity in video OF diagrams, respectively, are preserved. The resulting intensity-weighted velocity diagram is shown in Figure \ref{fig:OF}.

\begin{figure*}[t]
\centering
\includegraphics[height=3.8cm]{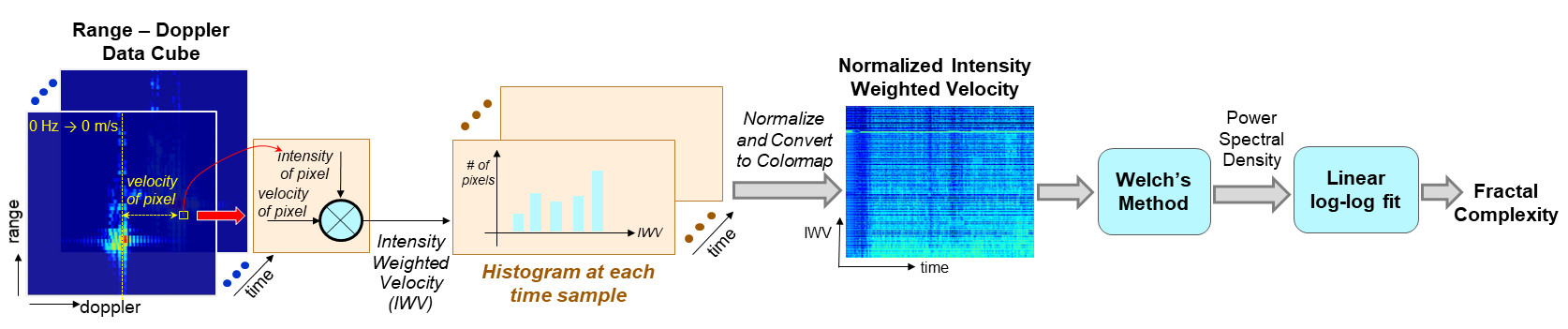}
  \caption{Block diagram showing computation of IWV diagrams and fractal complexity for RF data.}
  \label{fig:iwv}
\end{figure*}

\subsubsection{Fractal Complexity of RF Data}

The fractal complexity at each IWV is calculated by first finding the power spectral density (PSD) using Welch's Method \cite{1161901}, which essentially computes the Fourier transform of the IWV diagram versus time, as shown in Figure \ref{fig:iwv}.  This results in a magnitude matrix $M(j,f)$ where $(j,f)$ are velocity and frequency bins, respectively.  The fitting parameter for the fractal complexity, $\beta(j)$, is related to the magnitude at each velocity bin $j$ as
\begin{equation}
    M(j,f) = \frac{a}{\big| f \big|^{\beta}}
    \longrightarrow ln(M) = ln(a) - \beta ln{\big| f \big|},
\end{equation}
where $a$ is an amplitude fitting variable \cite{Evie7}.  
A simple linear fit is then performed on $ln(f)$ versus $ln(M)$, where $\beta$ relates to the slope and $ln(a)$ is the intercept on a log–log plot.  If $M(j,f)$ is integrated over $j$, an overall velocity spectrum can be obtained, which, after fitting, yields fractal complexity parameter, $\bar{\beta}$. Note that $\bar{\beta}$ is inversely related to fractal complexity, so that a lower $\bar{\beta}$ implies greater information. 

\subsubsection{Comparison of ASL with Daily Activities}

Because ASL is a language for interpersonal communication, it would be expected that it contain a greater amount of information than purely kinematic signals, such as obtained from daily activities.  This expectation can be verified through comparison of the fractal complexity of ASL with that of daily activities.  

The fractal complexity for ASL was computed using the sentence sequences acquired under Experiment 2, which yielded a total of 30 samples of 8 seconds duration each.  Daily activity data was acquired from two hearing participants enacting eight different daily activities: 1) building legos, 2) playing chess, 3) cooking, 4) painting, 5) eating, 6) vacuuming, 7) folding laundry, and 8) ironing clothes.  The test area was equipped with required props so that activities were conducted just as would be done were the participants at home.   Data was acquired for 10 minutes to ensure participants were moving about as naturally as possible. This data was then cropped into 8 second, non-overlapping segments to yield 30 samples per activity, or a total of 240 samples.  The average   $\bar{\beta}$ values calculated for ASL and daily activities using video and RF sensors are given in Table \ref{tab:betabar}.   For both sensors, the $\bar{\beta}$ value for ASL is less than that for daily activities. Thus, both RF sensors and video show that ASL signing communicates greater information than daily activities, and underscores the importance of not merely equating ASL with gesturing.

\begin{table}[h]
\centering
\includegraphics[height=1.5cm]{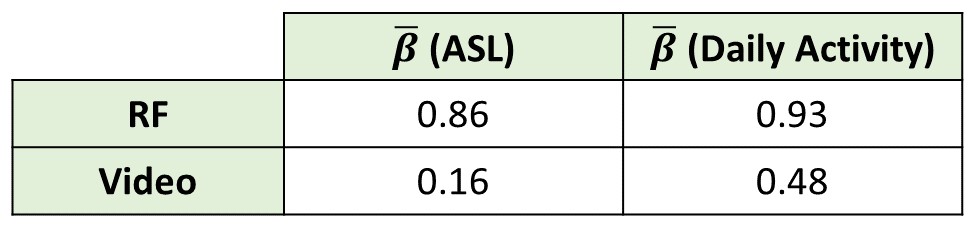}
  \caption{ $\bar{\beta}$ values for ASL and daily activities.}
  \label{tab:betabar}
\end{table}
\vspace{-3mm}
\subsection{Imitation versus Native ASL Signing}

Perceptual studies indicate that signers and non-signers differ drastically in their perception of rapidly changing visual stimuli \cite{Bosworth}. It can take learners of sign language \textit{at least 3 years} to produce signs in a manner that is perceived as fluent by native signers \cite{beal2018hearing}.

\subsubsection{Comparison of Native ASL and Imitation Signing Data}
It is significant that the differences in signing between native and imitation signers enumerated in sign language research literature can be revealed through visual observation and quantitative analysis of RF ASL data.

ASL is a fluid language that minimizes exertion.  But imitation signers are often hesitant or awkward, failing to replicate temporal tempo of signing.  Other errors of imitation signers include 
\begin{itemize}
    \item replicating signs with an incorrect number of repetitions; e.g. swiping fingers or moving hands back and forth too many times (e.g. \textsc{knife}, \textsc{write}, \textsc{car} or \textsc{walk}).
    \item exaggerating movements along inaccurate trajectories; e.g. thrusting arm forward rather than pointing via pivoting about the elbow when signing \textsc{you} or moving entire arm to clasp fingers rather than rotating hand about the wrist 
    \textsc{friend}
    This causes phonological and lexical errors. When signing \textsc{mountain}, imitation signers move their hands over a greater distance, with a different tempo, and open fingers slower than would a native signer; thus, the prosodic and pragmatic levels of signing are represented incorrectly.
    \item making gross motion errors; e.g.  shaking hands forwards/backwards rather than side to side or pointing rather than grabbing wrist (\textsc{earthquake}), shaking only one hand (\textsc{engineer}), or shifting entire body to point fingers over shoulder (\textsc{come}).
\end{itemize}

It is important to note that non-signers are not able to copy-sign connected discourse in sign language, as the speed and variety of sentence and narrative-level stimuli in sign language exceeds non-signer threshold of temporal resolution for non-signers \cite{Malaia_spatial}. While this much has been clear to all researchers (across current literature, only one study, \cite{fang2017deepasl}, attempted to have non-signers copy sentences - and  required an intense 3-hour training to accomplish this), it is worth noting that connected signing motion contains syntactic, prosodic, and pragmatic information that is absent in individual sign production \cite {malaia2013kinematic, malaia2012kinematic}.  This detrimentally affects the spatiotemporal distribution of the information-bearing signal that is characteristic of natural languages \cite{malaia2018information}.
In sum, copysigning by non-signers results in motor production data that does not approximate sign language production, but is severely distorted from the spectrotemporal perspective of sign language representation, and  contains linguistic errors that pervade phonological, semantic, syntactic, prosodic, and pragmatic representation. Copysigning data can only be used to study the trajectory of learning sign language – not sign language per se.

\subsubsection{Identification of Native ASL versus Imitation Signing}
Machine learning can be used to distinguish between native versus imitation signing.  The feature space spanned by native and imitation RF data can be compared using the PCA and T-distributed Stochastic Neighbor Embedding (t-SNE) algorithms, as shown in Figure 8. The t-SNE algorithm converts similarities between data points into joint probabilities and tries to minimize the Kullback-Leibler divergence between the joint probabilities. To remove any bias due to sample size, 180 samples were randomly selected from the imitation data set, an equal number to the native signing samples utilized.  It is striking that although there there is overlap between the two datasets, the variance of the imitation signers' data is greater than that of the native ASL signers.  Moreover, the centroids of each dataset are clearly shifted relative to one another, indicating that imitation signing is not truly representative of native ASL signing.  

\begin{figure}[t]
\centering
\includegraphics[height=4.5cm]{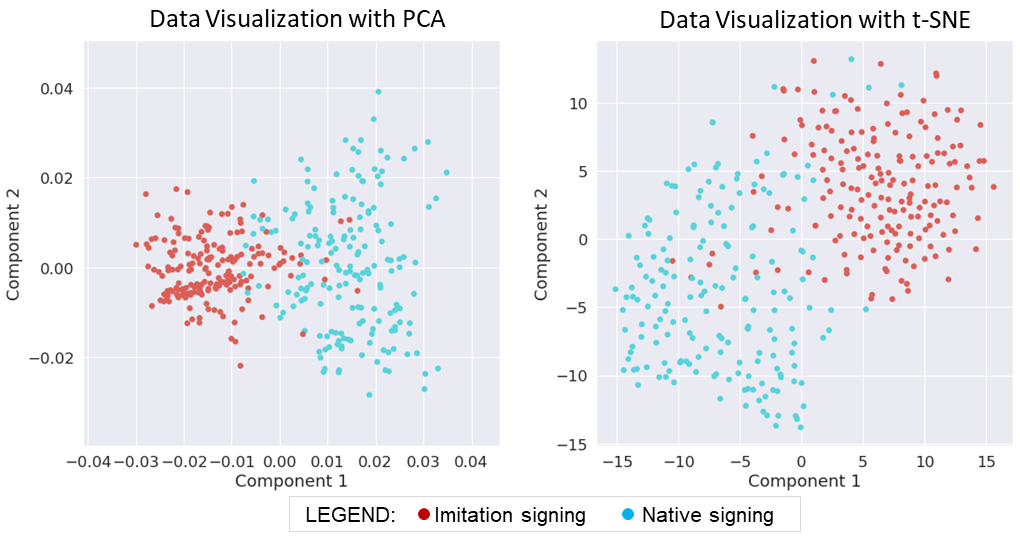}
  \caption{Comparison of RF data from native ASL users and imitation signers using PCA and tSNE.}
  \label{fig:tSNE}
\end{figure}

The difference between imitation and native ASL data can be further quantified by using a Suppor Vector Machine (SVM) classifier with Radial Basis Function (RBF) to explicitly classify imitation versus native signing data.   To mitigate class imbalance between the datasets, the synthetic minority  \cite{SMOTE2} technique (SMOTE) \cite{SMOTE1} was applied. SMOTE equalizes class data by oversampling minority classes with “synthetic” samples. 
SVM was applied on the features generated via PCA using five-fold cross validation.  We found that native and imitation signing could be distinguished with 99\% accuracy. 

These results underscore the need to test ASL recognition algorithms on native ASL data provided by Deaf participants, rather than rely upon more easily acquired imitation signing data from hearing participants via copysigning.

\section{ASL Recognition Using RF $\mu$D Signatures}
In this section, the ability of RF sensors to distinguish ASL signs is demonstrated through $\mu$D signature classification.  

\subsection{Handcrafted Features}
First, a wide range of features, drawn from four principal groupings, are extracted from the $\mu$D signatures:

\subsubsection{Envelope Features}
Envelope features have been shown to be significant physical features \cite{Gurbuz2015,Kim2015_Physical} of the $\mu$D signature as they describe the outline and peak response of the signature.  In this work, the maximum, minimum and mean value of upper and lower envelopes, as well as the difference between the average values of the upper and lower envelope are extracted using the percentile technique \cite{VanDorpGroen2008}.  


 \begin{figure*}[t!]
\centering
\includegraphics[width=17.5cm]{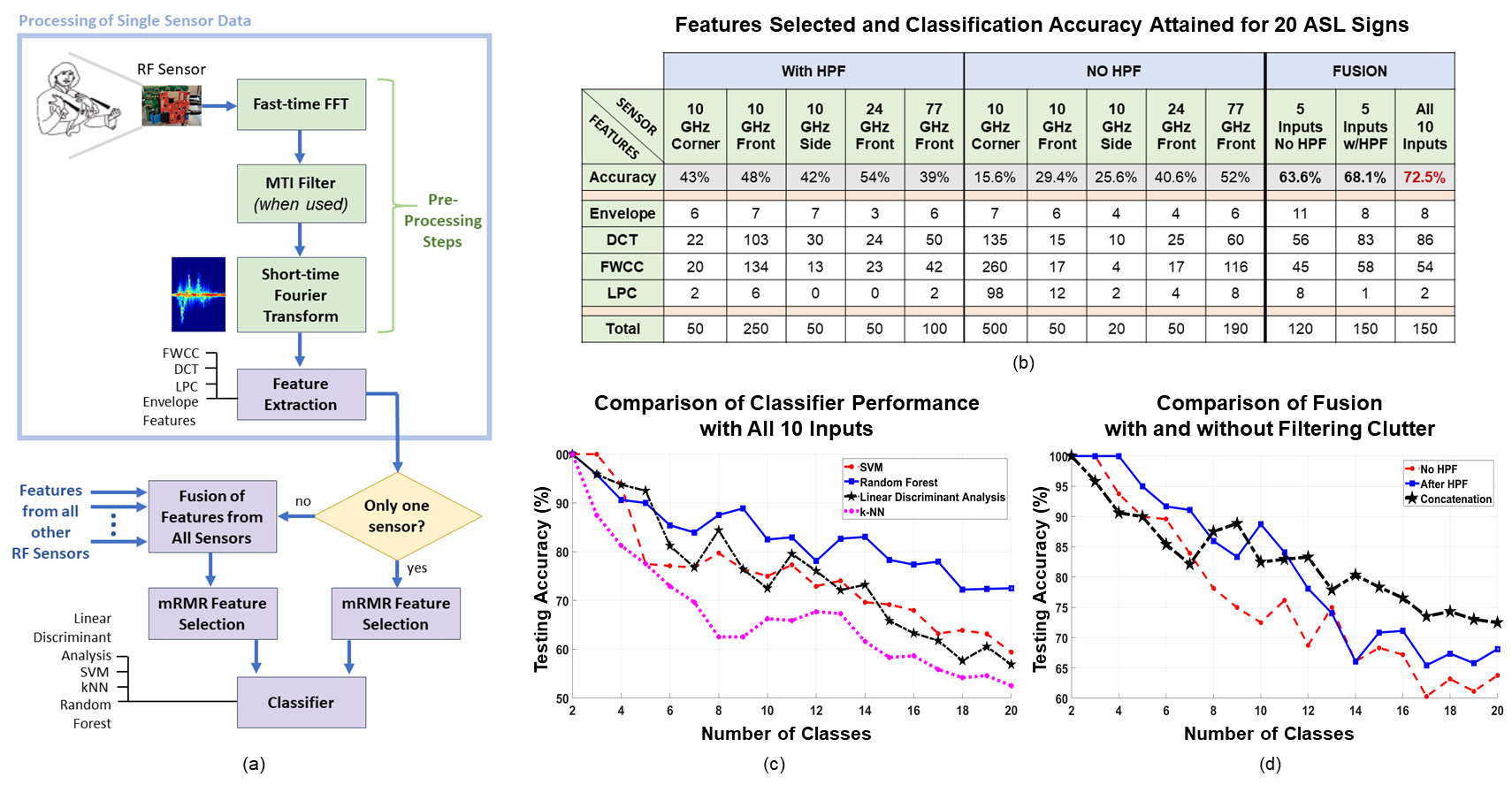}
  \caption{(a) Block diagram of ASL recognition approach using multi-frequency RF sensor network; (b) Features selected by MRMR algorithm and resulting accuracy with random forest classifier; (c) Comparison of performance given by different classifiers; (d) Comparison of performance achieved with and without filtering out clutter.}
  \label{fig:confmatrix}
\end{figure*}

\subsubsection{DCT Coefficients}
The Discrete Cosine Transform (DCT) represents a signal as the sum of sinusoidal functions oscillating at different frequencies.  We extracted 500 2-D DCT coefficients \cite{MolchanovDCT2011} from each $\mu$D signature and vectorized them.

\subsubsection{FWCC Features}
Frequency-warped cepstral coefficients (FWCC) are inspired from mel-frequency cepstral coefficients, common in speech processing, but whose filter bank is optimized to RF data using genetic algorithms \cite{ErolFWCC2018}.

FWCC features are found by first initializing a filter bank that contains $M$ triangular filters $h_{m}(k)$. Then, the log-energy output of each filter, $E_{m}$, is computed by applying the filter on the spectrogram ($S$) as follows:
\begin{equation}
    E_{(n,m)}=log \Big( \sum_{k=0}^{N-1} S(n,k) h_{m}(k) \Big)
\end{equation}
for $n=1,2, ... T$, where $N$ is the length of the filter and $T$ is time index of the spectrogram. We then compute the cepstral coefficients $C$ by taking the DCT of the filter outputs:
\begin{equation}
    C(j,n) =  \sum_{m=1}^{M} E_{(n,m)}  cos\Big(j(m-\frac{1}{2})\frac{\pi}{M} \Big)
\end{equation}
for $j=1,2, ... J$, where $J$ is the number of cepstral coefficients.  The feature vector, $C_{i}$, for $i^{th}$ sample is then obtained by vectorizing the resulting matrix $C(j,n)$. It should be noticed that the resulting coefficients depend on the parameters that we used while creating the filters; namely, its starting, maximum and end points.  The filter bank itself is optimized using a genetic algorithm, where the length of a chromosome is equal to three times the total number of filters, $3M$, since we have three parameters to define a filter. The classification performance of a random forest classifier with 100 decision trees is employed as the fitness function.

\subsubsection{LPC Coefficients}
Linear predictive coding (LPC) $\mu$D \cite{YKim_LPC2014} computes the coefficients of a forward linear predictor by minimizing the prediction error in a least squares sense. In this work, 100 LPC coefficients were computed from each $\mu$D spectrogram by minimizing the residual error, $e[n]$ in
\begin{equation}
    s[n]=\sum_{k=1}^{p} a_{k} s[n-k] + e[n],
\end{equation}
where $\{a_k\}$ are $p^{th}$ order linear predictor coefficients and $s$ is the spectrogram, first on the columns and then the rows.

 \subsection{Feature Selection and Classification Results}
 Due to the curse of dimensionality, an effective subset of features is oftentimes more  beneficial in classification than blindly using all possible features.  While there are many available methods for feature selection, it has been shown \cite{Tekeli2016} that for $\mu D$ classification the minimum redundancy maximum relevance (mRMR) algorithm \cite{PengMRMR2005} yields the best performance without having any dependency on the specific classifier used, while also having better generalization capabilities. A total of 932 features are initially extracted for each RF sensor in the network: 7 envelope features, 500 DCT coefficients, 325 FWCC features, and 100 LPC coefficients.  For all sensors, features were extracted from the spectrogram both with and without high pass filtering (HPF). This is because while HPF removes stationary clutter, there is also the potential for important low-frequency information to be lost. Next, the mRMR algorithm was applied to select an optimal subset of features from each sensor.  The number of features selected was varied between 20 and 250, while four different classifiers considered to evaluate performance:  support vector machines (SVM), k-nearest neighbors (kNN), linear discriminant analysis (LDA) using random subspace gradient boosting, and random forest classifier (RFC).  During classification, 75\% of the data was used for training, while 25\% was used for testing. An overview of the proposed approach is given in Figure \ref{fig:confmatrix}(a).

 Table \ref{fig:confmatrix}(b) shows the selected features and accuracy obtained for classification of 20 ASL signs  using RFC, as that was the classifier that offered the best overall performance.  Notice that significant numbers of all types of features were selected, with the exception of LPC, which does not appear to be very effective towards ASL recognition.  In contrast, features that reflect the decomposition of the signature into different frequency components, such as DCT and FWCC, are strongly preferred from an information theoretic perspective.  
 
 Better performance is achieved at the higher transmit frequencies, with the accuracy of the 24 GHz sensor closely followed by that of the 77 GHz sensor.  Notice that there is a great benefit to not using a HPF on 77 GHz data, for which the accuracy without the HPF is increased by 13\% whereas other sensors benefit from the filtering.  The ASL recognition accuracy can be greatly increased by fusing features form all inputs, and performing a second round of feature selection with mRMR.  This yields a classification accuracy of 72\% for 20 ASL classes - about a 15\% - 30\% improvement over the results obtained with just a single sensor. A classification accuracy of 95\% is attained for recognition of 5 ASL signs.  The results obtained for all four classifiers as a function of the total number of signs discriminated is given in Figure \ref{fig:confmatrix}(c), while the impact of HPF is compared in Figure \ref{fig:confmatrix}(d).


\section{Discussion}
It should be emphasized that the results presented in Figure \ref{fig:confmatrix} used data obtained from native ASL signers during both training and testing.  While imitation data, which is acquired by  hearing  non-signers via copysigning, is easier to collect and can provide a large data set quickly, imitation data differ substantially from native signing data, as shown in Section IV-C. 
Consequently, machine learning algorithms should be benchmarked using test data from native ASL signers, while the use of imitation data during training is not effective for the classification of native ASL signing data.

Moreover, it is difficult to make a direct comparison between the results of this RF sensing study versus those attained from alternative modalities because published projects (e.g. \cite{C1,C2,C3,fang2017deepasl,SignFi2018,WiSign2017}) tended to use imitation data for both model training \textit{and} testing.  Thus, there are discrepancies both in terms of amount of available training data and statistical properties of the test set. A  large multimodal  database  of  connected  native  signing would be needed to draw meaningful conclusions for technology and algorithm comparison.

The trade-off in data quantity versus fidelity has been very apparent in our on-going work relating to the development of domain adaptation algorithms that would facilitate the exploitation of imitation signing data for the classification of native ASL signing data.  Preliminary results indicate that this approach can be one way to increase the amount of suitable training data available and facilitate the design of deep neural networks that offer a substantial increase in recognition accuracy of dynamic signing using RF sensors.
  
\section{Conclusion}
This paper provides an in-depth, multi-disciplinary perspective on ASL recognition with RF sensors, and is (to the best of our knowledge) the first work on RF sensing of native ASL signing.  We demonstrate that RF sensing can be a useful tool in linguistic analysis, capturing properties such as co-articulation and exhibiting greater fractal complexity (i.e. information content) than daily activities.  Significantly, based on linguistics and machine learning, we show that RF sensors can reveal a discernible difference between imitation signing and native signing. Frequency warped cepstral coefficients (FWCC) are optimized for ASL using genetic algorithms, and in conjunction with Discrete Cosine Transform (DCT) coefficients, and envelope features,  used to classify up to 20 ASL signs.  Using the minimum  redundancy  maximum  relevance (mRMR)  algorithm, an optimal subset of 150 features are selected and input to a random forest classifier to achieve 95\% recognition accuracy for 5 signs and 72\% accuracy for 20 signs. These results demonstrate the potential of RF sensing to provide non-contact ASL recognition capabilities in support of ASL-sensitive smart environments, while remaining effective in the dark and protecting user privacy.


%

\section*{Acknowledgment}
The authors would like to thank Dr. Caroline Kobek-Pezzarossi from Gallaudet University, Washington D.C. and Dr. Dennis Gilliam from AIDB for their support of this research. This work was funded in part by the National Science Foundation (NSF) Cyber-Physical Systems (CPS) Program Awards \#1932547 and \#1931861, NSF Integrative Strategies for Understanding Neural and Cognitive Systems  (NCS) Program Award \#1734938, as well as the University of Alabama ECE and CS Departments.  Human studies research was conducted under UA Institutional Review Board (IRB) Protocol \#18-06-1271.





\bibliographystyle{IEEEtran}
\bibliography{Evie-refs}
\end{document}